\input{aipcheck}

\documentclass[
    ,final            
  ]
  {aipproc}
  
\usepackage{amssymb}

\layoutstyle{8x11double}

\begin{document}

\title{QCD Thermodynamics at Intermediate Coupling}


\author{Nan Su}{
  address={Frankfurt Institute for Advanced Studies, Ruth-Moufang-Str. 1, D-60438 Frankfurt am Main, Germany}
}



\begin{abstract}

The weak-coupling expansion of the QCD free energy is known to order $g_s^6\log{g_s}$, however, the resulting series is poorly convergent at phenomenologically relevant temperatures. In this proceedings, I discuss hard-thermal-loop perturbation theory (HTLpt) which is a gauge-invariant reorganization of the perturbative expansion for gauge theories. I review a recent NNLO HTLpt calculation of QCD thermodynamic functions. I show that the NNLO HTLpt results are consistent with lattice data down to temperatures $\sim2T_c$.

\end{abstract}

\maketitle


\section{Introduction}

The current generation of ultrarelativistic heavy-ion collision experiments should exceed the energy density necessary for the formation of a quark-gluon plasma. Initial temperatures of RHIC are up to twice the QCD critical temperature, $T_c \sim 170$ MeV. The strong coupling constant at these initial temperatures is approximately $g_s \sim 2$ or $\alpha_s = g_s^2/4\pi \sim 0.3$, which is some intermediate value, neither infinitesimally small nor infinitely large. Theoretically, one expected that this state of matter could be described in terms of weakly interacting quasiparticles; however, data from RHIC suggested that the state of matter created there behaved more like a strongly coupled fluid with a small viscosity~\cite{rhicexperiment}. This has inspired work on strongly-coupled formalisms. However, some observables such as jet quenching~\cite{jet} and elliptic flow~\cite{elliptic} can also be described using perturbative methods and so it is difficult to judge whether the plasma is strongly or weakly coupled based only on RHIC data. The initial temperatures of the upcoming experiments at LHC are expected up to $4-6\,T_c$ and due to asymptotic freedom of QCD, this corresponds to a smaller coupling constant. A key question is then whether the matter generated can be described in terms of weakly interacting quasiparticles at these higher temperatures.

The weak-coupling expansion of the QCD free energy is known up to order $g_s^6\log g_s$~\cite{npert,bn}. Unfortunately, the resulting series shows no sign of convergence at phenomenologically relevant temperatures. There are several ways of reorganizing the perturbative series at finite temperature~\cite{reviews} and they are all based on a quasiparticle picture where one is perturbing about an ideal gas of massive quasiparticles, rather than that of massless particles. In the following I will discuss recent advances in the application of hard-thermal-loop perturbation theory (HTLpt).

\section{Hard-thermal-loop perturbation theory}

Hard-thermal-loop perturbation theory is a gauge-invariant extension of screened perturbation theory~\cite{spt}. The basic idea of the technique is to add and subtract an effective mass term from the bare Lagrangian, and to associate the added piece with the free Lagrangian and the subtracted piece with the interactions. In gauge theories, however, simply adding and subtracting a local mass term violates gauge invariance~\cite{gaugebreak}. Instead one adds and subtracts an HTL improvement term, which dresses the propagators and vertices self-consistently so that the reorganization is manifestly gauge invariant~\cite{Braaten:1991gm}. HTLpt has recently been pushed to NNLO and the details of the formalism and calculations are presented in Refs.~\cite{qed,ym,qcd1}. Here only a few selected results from QCD~\cite{qcd1} are reviewed.

\begin{figure}[t]
\includegraphics[width=6.6cm]{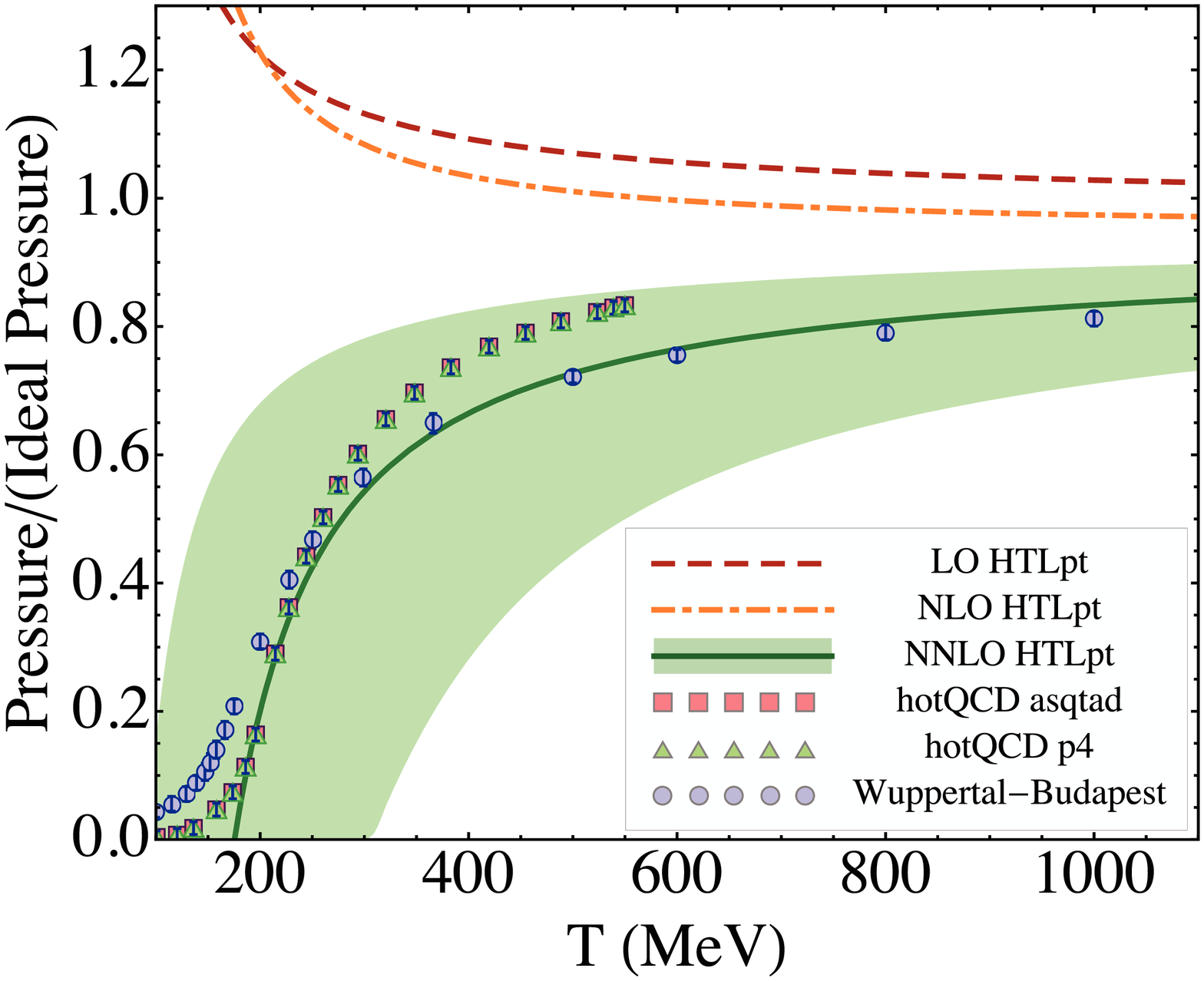}
\hspace{2mm}
\includegraphics[width=6.6cm]{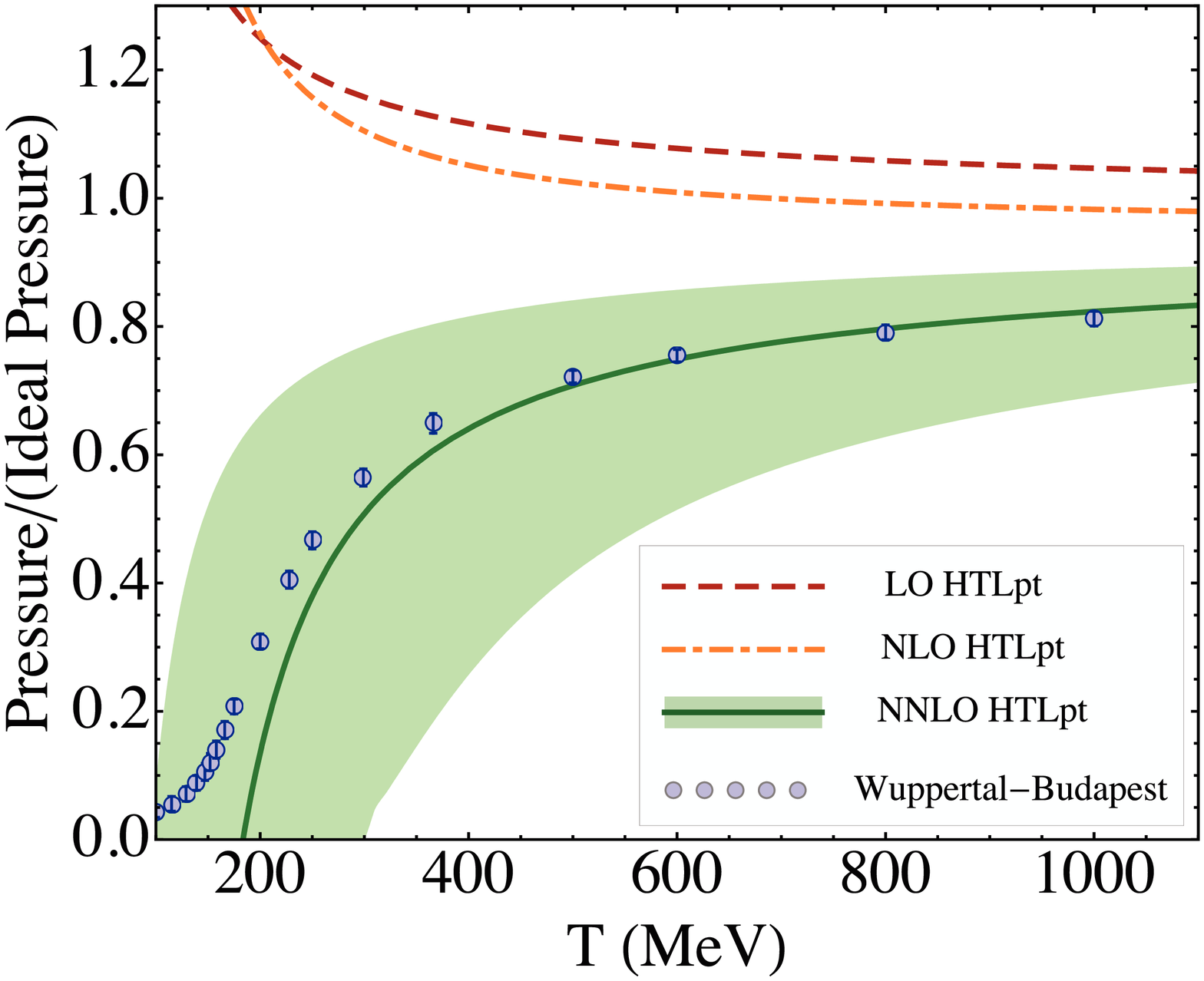}
\caption{Comparison of LO, NLO, and NNLO predictions for the scaled pressure for $N_f=3$ (left panel) and $N_f=4$ (right panel) with lattice data from Bazavov et al.~\cite{hotqcd} and Borsanyi et al. \cite{wupp-buda}. See main text for details.}
\label{pressure}
\end{figure}

With rescaled dimensionless parameters $\hat m_{D/q} = m_{D/q} /(2 \pi T)$ and $\hat\mu = \mu /(2 \pi T)$, the renormalized NNLO thermodynamic potential for QCD with $N_f$ flavors and $N_c$ colors reads
\begin{eqnarray}\nonumber
{\Omega_{\rm NNLO}\over{\cal F}_{\rm ideal}}
\!\!&=&\!\!
1+{7\over4}{d_F\over d_A}
-{15\over4}\hat{m}_D^3
+{c_A\alpha_s\over3\pi}\left[
-{15\over4}
+{45\over2}\hat{m}_D
\right. \\ && \left. \nonumber
\hspace{-18mm}
-\;{135\over2}\hat{m}^2_D
-{495\over4}\left(\log{\hat\mu \over 2}+{5\over22}+\gamma_E\right)\hat m_D^3
\right]+{s_F\alpha_s\over\pi}
\\ && \nonumber
\hspace{-17.5mm}
\times\left[-{25\over8}
+{15\over2}\hat{m}_D
+15\left(
\log{\hat\mu \over 2}-{1\over2}+\gamma_E+2\log2\right)\hat m_D^3
\right. \\ && \left. \nonumber
\hspace{-18mm}
-\;90\hat{m}^2_q\hat{m}_D\right]
+\left({c_A\alpha_s\over3\pi}\right)^2\left[{45\over4}{1\over\hat{m}_D}
-{165\over8}\left(\log{\hat{\mu}\over2}
\right.\right.\\ &&\left.\left. \nonumber
\hspace{-18.5mm}
-\;{72\over11}\log{\hat{m}_D}
-{84\over55}
-{6\over11}\gamma_E
-{74\over11}{\zeta^{\prime}(-1)\over\zeta(-1)}
+{19\over11}{\zeta^{\prime}(-3)\over\zeta(-3)}
\right)
\right. \\ && \left. \nonumber
\hspace{-18mm}
+\;{1485\over4}
\left(
\log{\hat{\mu}\over2}
-{79\over44}+\gamma_E+\log2-{\pi^2\over11}
\right)\hat{m}_D
\right]
\\&&\nonumber
\hspace{-17.5mm}
+\left({c_A\alpha_s\over3\pi}\right)\left({s_F\alpha_s\over\pi}\right)
\left[
{15\over2}{1\over\hat{m}_D}
-{235\over16}\left(\log{\hat{\mu}\over2}
\right.\right.\\ &&\nonumber\left.\left.
\hspace{-18mm}
-\;{144\over47}\log{\hat{m}_D}
-{24\over47}\gamma_E
+{319\over940}+{111\over235}\log2
-{74\over47}{\zeta^{\prime}(-1)\over\zeta(-1)}
\right.\right.\\ &&\nonumber\left.\left.
\hspace{-18mm}
+\;{1\over47}{\zeta^{\prime}(-3)\over\zeta(-3)}
\right)
+{315\over4}\left(\log{\hat{\mu}\over2}-{8\over7}\log2+\gamma_E
\right.\right.\\ &&\nonumber\left.\left.
\hspace{-18mm}
+\;{9\over14}
\right)\hat{m}_D
+90{\hat{m}_q^2\over\hat{m}_D}
\right]
+\left({s_F\alpha_s\over\pi}\right)^2
\left[{5\over4}{1\over\hat{m}_D}
+{25\over12}\left(
\log{\hat{\mu}\over2}
\right.\right.\\ &&\nonumber\left.\left.
\hspace{-18mm}
+\;{1\over20}+{3\over5}\gamma_E-{66\over25}\log2
+{4\over5}{\zeta^{\prime}(-1)\over\zeta(-1)}
-{2\over5}{\zeta^{\prime}(-3)\over\zeta(-3)}\right)
\right.\\ && \left. \nonumber
\hspace{-18mm}
-\;15\left(\log{\hat{\mu}\over2}
-{1\over2}+\gamma_E+2\log2
\right)\hat{m}_D
+30{\hat{m}_q^2\over\hat{m}_D}
\right]
\\ &&
\hspace{-18mm}
+\;s_{2F}\left({\alpha_s\over\pi}\right)^2\left[{15\over64}(35-32\log2)
-{45\over2}\hat{m}_D\right].
\label{Omega-NNLO}
\end{eqnarray}

In order to complete a calculation, a prescription is required to determine the mass parameters $m_D$ and $m_q$. Here the Debye mass is set to the mass parameter of three-dimensional electric QCD (EQCD)~\cite{bn}, i.e.~$m_D=m_E$. In Ref.~\cite{bn}, it was calculated to NLO giving
\begin{eqnarray}\nonumber
m_D^2\!\!&=&\!\!{4\pi\alpha_s\over3}T^2\left\{
c_A+s_F
+
{c_A^2\alpha_s\over3\pi}\left(
{5\over4}+{11\over2}\gamma_E
\right.\right. \\ && \left.\left. \nonumber
\hspace{-13mm}
+\;{11\over2}\log{\hat{\mu}\over2}\right)
+{c_As_F\alpha_s\over\pi}\left(
{3\over4}-{4\over3}\log2
+{7\over6} \gamma_E
+{7\over6} \log{{\hat{\mu}\over2}}\right)
\right.\\ &&\left.
\hspace{-13mm}
+\;{s_F^2\alpha_s\over\pi}\left(
{1\over3}-{4\over3}\log2-{2\over3}\gamma_E
-{2\over3}\log{{\hat{\mu}\over2}}
\right)
-{3\over2}{s_{2F}\alpha_s\over\pi}
\right\}.
\label{bnmass}
\end{eqnarray}
The quark mass is set to $m_q=0$.

In Fig.~\ref{pressure}, I show the scaled QCD pressure for $N_f=3$ (left panel) and $N_f=4$ (right panel) as a function of $T$. The results at LO, NLO, and NNLO use the BN mass given by Eq.~(\ref{bnmass}) as well as $m_q=0$. For the strong coupling constant $\alpha_s$, three-loop running~\cite{partdata} with $\Lambda_{\overline{\rm MS}}=344$ MeV~\cite{McNeile:2010ji} is used here. The bands correspond to varying the renormalization scale $\mu$ by a factor of 2 around $\mu = 2\pi T$ which are the central lines.

The lattice data from the Wuppertal-Budapest collaboration uses the stout action. 
Since their results show essentially no dependence on the lattice spacings (it is smaller than the statistical errors), they provide a continuum estimate by averaging
the trace anomaly measured using their two smallest lattice spacings corresponding to $N_\tau = 8$ and $N_\tau = 10$ \cite{wupp-buda}. Using standard lattice techniques, the continuum-estimated pressure is computed from an integral of the trace anomaly. The lattice data from the hotQCD collaboration are their $N_\tau = 8$ results using both the asqtad and p4 actions~\cite{hotqcd}. The hotQCD results have not been continuum extrapolated and the error bars correspond to only statistical errors and do not factor in the systematic error associated with the calculation which, for the pressure, is estimated by the hotQCD collaboration to be between 5 - 10\%.

\begin{figure}[t]
\includegraphics[width=6.6cm]{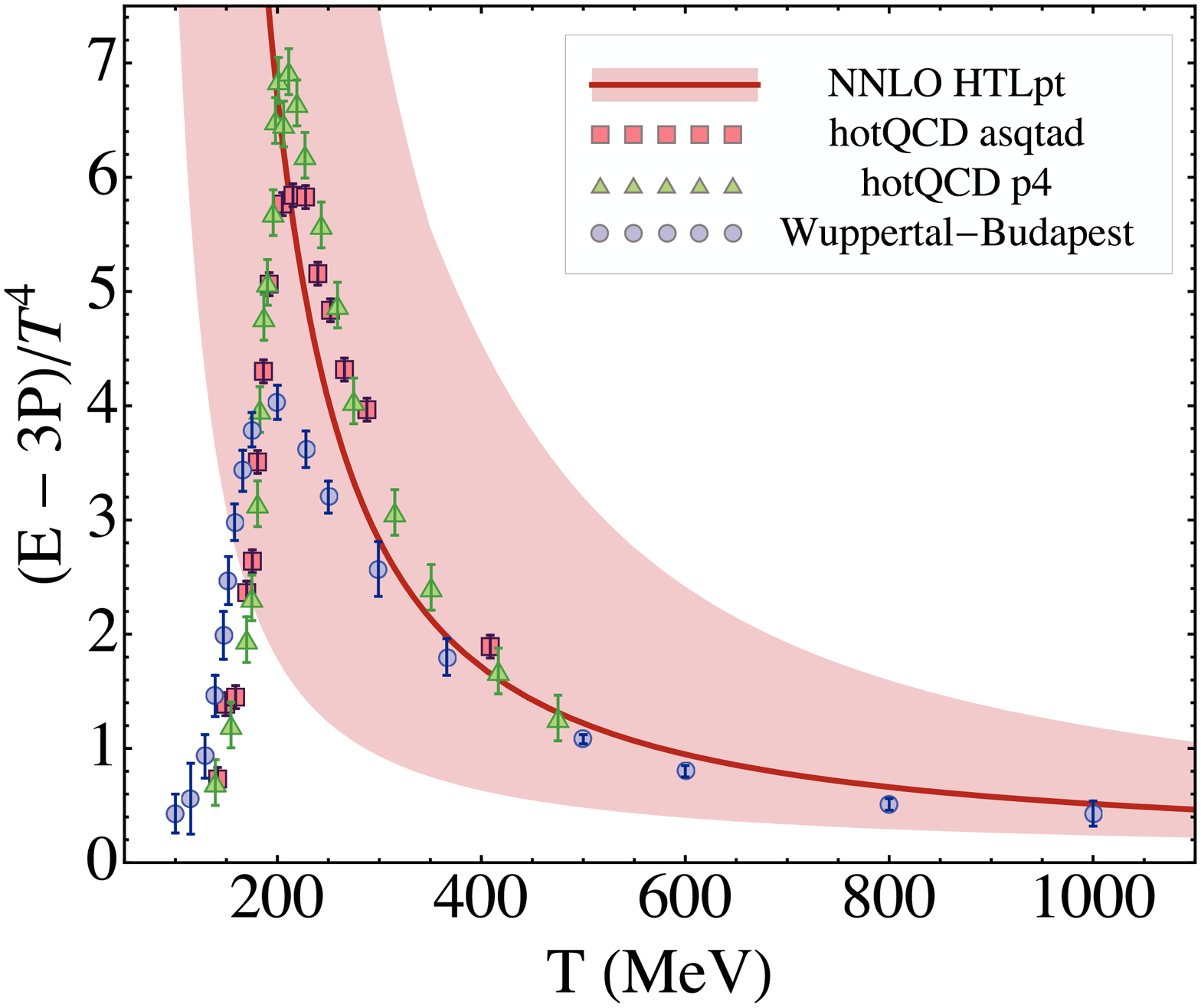}
\hspace{2mm}
\includegraphics[width=6.6cm]{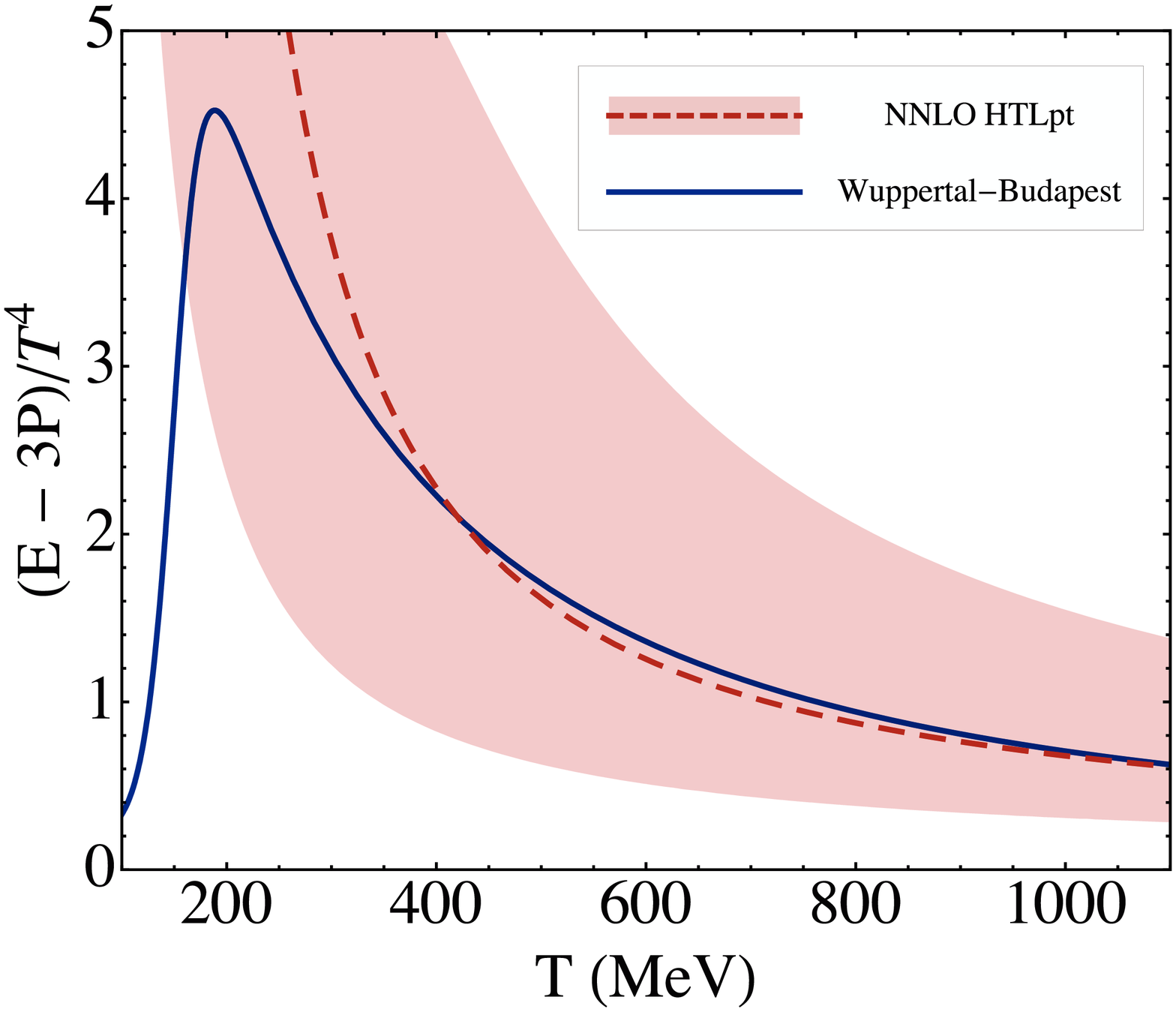}
\caption{Comparison of NNLO predictions for the scaled trace anomaly with $N_f=3$ (left panel) and $N_f=4$ fermions (right panel) lattice data from Bazavov et al.~\cite{hotqcd} and Borsanyi et al. \cite{wupp-buda}. See main text for details.
}
\label{trace2}
\end{figure}

As can be seen from Fig.~\ref{pressure} the successive HTLpt approximations represent an improvement over that of a naive weak-coupling expansion; however, as in the pure-glue case~\cite{ym}, the NNLO result represents a significant correction to the LO and NLO results. That being said the NNLO HTLpt result agrees quite well with the available lattice data down to temperatures on the order of $2\,T_c \sim 340$ MeV for both $N_f=3$ and $N_f=4$.

In Fig.~\ref{trace2}, I show the NNLO approximation to the scaled QCD trace anomaly as a function of $T$ for $N_f=3$ (left panel) and $N_f=4$ (right panel). The left panel shows data from both the Wuppertal-Budapest collaboration and the hotQCD collaboration taken from the same data sets displayed in Fig.~\ref{pressure}. In the case of the hotQCD, the results for the trace anomaly using the p4 action show large lattice size affects at all temperatures shown and the asqtad results for the trace anomaly show large lattice size effects for $T \sim 200$ MeV. The right panel displays a parameterization (solid blue curve) of the trace anomaly for $N_f=4$ published by the Wuppertal-Budapest collaboration \cite{wupp-buda} since the individual data points were not published. Both panels show very good agreement with the available lattice data down to temperatures on the order of $T \sim 2\,T_c$. 
Note that due to the massless quark description in HTLpt, deviations are expected from the lattice data for $T\lesssim414$ MeV.

\section{Conclusions and outlook}

In this proceedings, I briefly reviewed recent NNLO results for the QCD thermodynamics using HTLpt. From comparison with lattice data for $N_f \in \{3,4\}$, it has been shown that HTLpt is consistent with available lattice data down to $T\sim2\,T_c$ for the pressure and the trace anomaly.

In closing, I emphasize that HTLpt provides a gauge invariant reorganization of perturbation theory for calculating static and dynamic quantities in thermal field theory. Given the good agreement with lattice data for thermodynamics, it would be interesting to apply HTLpt to the calculation of  real-time quantities at temperatures that are relevant for LHC.


\emph {This work was done in collaboration with Jens O. Andersen, Lars E. Leganger and Michael Strickland. The author was supported by the Frankfurt International Graduate School for Science and Helmholtz Graduate School for Hadron and Ion Research.}



\bibliographystyle{aipproc}   

\bibliography{sample}

\begin{thebibliography}{9}

\bibitem{rhicexperiment}
  J.~Adams {\it et al.}
  Nucl.\ Phys.\  A {\bf 757}, 102 (2005);
  K.~Adcox {\it et al.}, {\em ibid.}, 184 (2005);
  I.~Arsene {\it et al.}, {\em ibid.}, 1 (2005);
  B.~B.~Back {\it et al.}, {\em ibid.}, 28 (2005);
  M.~Gyulassy and L.~McLerran,
  Nucl.\ Phys.\  A {\bf 750}, 30 (2005).

\bibitem{jet}
  G.~Y.~Qin, J.~Ruppert, C.~Gale, S.~Jeon, G.~D.~Moore and M.~G.~Mustafa,
  Phys.\ Rev.\ Lett.\ {\bf 100} (2008) 072301;
  G.~Y.~Qin and A.~Majumder,
  arXiv:0910.3016 [hep-ph].

\bibitem{elliptic}
  Z.~Xu C.~Greiner, and H.~Stocker
  Phys.\ Rev.\ Lett.\  {\bf 100}, 172301 (2008).

\bibitem{hotqcd}
  A.~Bazavov {\it et al.},
  Phys.\ Rev.\  D {\bf  80}, 014504 (2009).

\bibitem{wupp-buda}
  S.~Borsanyi {\it et al.},
  arXiv:1007.2580 [hep-lat].

\bibitem{npert}
  E.~V.~Shuryak,
  Sov.\ Phys.\ JETP {\bf 47} (1978) 212
  [Zh.\ Eksp.\ Teor.\ Fiz.\ {\bf 74}, 408 (1978) ];
  J.~I. Kapusta,
  Nucl.\ Phys. B {\bf 148} (1979) 461;
  T.~Toimela,
  Int.\ J.\ Theor.\ Phys.\ {\bf 24}, 901(1985)
  [Erratum-{\em ibid.}\ {\bf 26}, 1021 (1987)];
  P.~B.~Arnold and C.~X.~Zhai,
  Phys.\ Rev.\ D {\bf 50}, 7603 (1994);
  Phys.\ Rev.\ D {\bf 51}, 1906 (1995);
  Phys.\ Rev.\ D {\bf 53}, 3421 (1996);
  C.~X.~Zhai and B.~Kastening,
  Phys.\ Rev.\ D {\bf 52}, 7232 (1995);
  K.~Kajantie, M.~Laine, K.~Rummukainen and Y.~Schroder,
  Phys.\ Rev.\ D {\bf 67}, 105008 (2003).

\bibitem{bn}
  E.~Braaten and A.~Nieto,
  Phys.\ Rev.\ Lett.\ {\bf 76}, 1417 (1996);
  Phys.\ Rev.\ D {\bf 53}, 3421 (1996).

\bibitem{reviews}
  J.~P.~Blaizot, E.~Iancu and A.~Rebhan,
  In {\it Hwa, R.C. (ed.) et al.: Quark gluon plasma},  60-122, (2003);
  U. Kraemmer and A. Rebhan,
  Rept. Prog. Phys. {\bf 67}, 351 (2004);
  J.~O.~Andersen and M.~Strickland,
  Annals Phys. {\bf 317}, 281 (2005).

\bibitem{spt}
  F.~Karsch, A.~Patk\'os, and P.~Petreczky,
  Phys.~Lett.~{\bf B401}, 69 (1997);
  S.~Chiku and T.~Hatsuda,
  Phys.\ Rev.\ {\bf D58} (1998) 076001;
  J.~O.~Andersen, E.~Braaten and M.~Strickland,
  Phys.\ Rev.\ {\bf D63}, 105008 (2001);
  J.~O.~Andersen and L.~Kyllingstad,
  Phys.\ Rev. {\bf D78}, 076008 (2008).

\bibitem{gaugebreak}
  W.~Buchm\"uller and O.~Philipsen,
  Nucl.\ Phys.\ B {\bf 443}, 47  (1995);
  G.~Alexanian and V.~P.~Nair,
  Phys.\ Lett.\ B {\bf 352}, 435 (1995).

\bibitem{Braaten:1991gm}
  E.~Braaten and R.~D.~Pisarski,
  Phys.\ Rev.\ D {\bf 45}, R1827 (1992).

\bibitem{qed}
  J.~O.~Andersen, M.~Strickland and N.~Su,
  Phys.\ Rev.\  D {\bf 80}, 085015 (2009).

\bibitem{ym}
  J.~O.~Andersen, M.~Strickland and N.~Su,
  Phys.\ Rev.\ Lett.\  {\bf 104}, 122003 (2010);
  JHEP {\bf 1008}, 113 (2010).

\bibitem{qcd1}
  J.~O.~Andersen, L.~E.~Leganger, M.~Strickland and N.~Su,
  arXiv:1009.4644 [hep-ph].

\bibitem{partdata}
 C. Amsler {\it et al.} (Particle Data Group), Physics Letters B {\bf 667}, 1 (2008).

\bibitem{McNeile:2010ji}
  C.~McNeile, C.~T.~H.~Davies, E.~Follana, K.~Hornbostel and G.~P.~Lepage,
  Phys.\ Rev.\  D {\bf 82}, 034512 (2010).


\end{thebibliography}

\IfFileExists{\jobname.bbl}{}
 {\typeout{}
  \typeout{******************************************}
  \typeout{** Please run "bibtex \jobname" to optain}
  \typeout{** the bibliography and then re-run LaTeX}
  \typeout{** twice to fix the references!}
  \typeout{******************************************}
  \typeout{}
 }


\end{document}